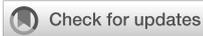

# The association between neighborhood obesogenic factors and prostate cancer risk and mortality: the Southern Community Cohort Study

Fekede Asefa Kumsa[1]*, Jay H. Fowke[2], Soheil Hashtarkhani[1], Brianna M. White[1], Martha J. Shrubsole[3] and Arash Shaban-Nejad[1]*

[1]Department of Pediatrics, College of Medicine, The University of Tennessee Health Science Center (UTHSC) - Oak Ridge National Laboratory (ORNL) Center for Biomedical Informatics, Memphis, TN, United States, [2]Department of Preventive Medicine, College of Medicine, University of Tennessee Health Science Center, Memphis, TN, United States, [3]Department of Medicine, Vanderbilt Epidemiology Center, Vanderbilt Ingram Cancer Center, Vanderbilt University Medical Center, Nashville, TN, United States

**Background:** Prostate cancer is one of the leading causes of cancer-related mortality among men in the United States. We examined the role of neighborhood obesogenic attributes on prostate cancer risk and mortality in the Southern Community Cohort Study (SCCS).

**Methods:** From the total of 34,166 SCCS male participants, 28,356 were included in the analysis. We assessed the relationship between neighborhood obesogenic factors [neighborhood socioeconomic status (nSES) and neighborhood obesogenic environment indices including the restaurant environment index, the retail food environment index, parks, recreational facilities, and businesses] and prostate cancer risk and mortality by controlling for individual-level factors using a multivariable Cox proportional hazards model. We further stratified prostate cancer risk analysis by race and body mass index (BMI).

**Results:** Median follow-up time was 133 months [interquartile range (IQR): 103, 152], and the mean age was 51.62 (SD: ± 8.42) years. There were 1,524 (5.37%) prostate cancer diagnoses and 98 (6.43%) prostate cancer deaths during follow-up. Compared to participants residing in the wealthiest quintile, those residing in the poorest quintile had a higher risk of prostate cancer (aHR = 1.32, 95% CI 1.12–1.57, $p$ = 0.001), particularly among non-obese men with a BMI < 30 (aHR = 1.46, 95% CI 1.07–1.98, $p$ = 0.016). The restaurant environment index was associated with a higher prostate cancer risk in overweight (BMI $\geq$ 25) White men (aHR = 3.37, 95% CI 1.04–10.94, $p$ = 0.043, quintile 1 vs. None). Obese Black individuals without any neighborhood recreational facilities had a 42% higher risk (aHR = 1.42, 95% CI 1.04–1.94, $p$ = 0.026) compared to those with any access. Compared to residents in the wealthiest quintile and most walkable area, those residing within the poorest quintile (aHR = 3.43, 95% CI 1.54–7.64, $p$ = 0.003) or the least walkable area (aHR = 3.45, 95% CI 1.22–9.78, $p$ = 0.020) had a higher risk of prostate cancer death.

**Conclusion:** Living in a lower-nSES area was associated with a higher prostate cancer risk, particularly among Black men. Restaurant and retail food environment indices were also associated with a higher prostate cancer risk, with stronger associations within overweight White individuals. Finally, residing in a low-SES neighborhood or the least walkable areas were associated with a higher risk of prostate cancer mortality.

KEYWORDS

prostate cancer, mortality, neighborhood obesogenic factors, neighborhood socioeconomic environments, built environments, the Southern Community Cohort Study

# Introduction

Prostate cancer remains the most common cancer diagnosis and a leading cause of cancer-related mortality among men in the United States (1). Established risk factors for prostate cancer include age (2) and family history (3), in addition to several genetic susceptibility markers (4). Non-genetic risk factors either at the individual level or as attributes of the neighborhood or built environment remain less well understood. Social determinants of health (SDoH) include neighborhood socioeconomic status (nSES), neighborhood healthcare access, and income inequality (5). However, several registry-based analyses found inconsistent relationships between indices of nSES, neighborhood deprivation, or segregation with prostate cancer risk, aggressiveness, or mortality (6–8).

Most prior studies have reported a consistent link between increasing obesity and prostate cancer mortality, and several studies report that obese men are more likely to be diagnosed with high-grade prostate cancer (9–11). However, population-based analyses rarely make explicit the environment in which individuals develop prostate cancer. Obesity is multifactorial (12) and may be affected by SDoH as mediated through the availability of nutritious foods or opportunities for a physically active lifestyle (13). Furthermore, the built environment may contribute to race differences in any obesity and prostate cancer analysis. A recent prospective analysis in the Multiethnic Cohort Study (MEC) based in California and Hawai'i reported that lower nSES was associated with lower overall and low-grade prostate cancer risk, with the strongest impact among foreign-born Latino men (14). Interestingly, the retail food environment index as an estimate of unhealthy-to-healthy food sources in the neighborhood and perhaps more directly linked to an obesogenic environment was not associated with prostate cancer incidence in the MEC.

Our goal is to investigate the social determinants of obesity and how these may be associated with prostate cancer outcomes in White and Black men. Our analyses include men living in the southeastern U.S. and participating in the Southern Community Cohort Study (SCCS). The SCCS recruitment included a substantial number of Black and White lower-income participants with comparable access to healthcare services. We prospectively investigate the role of the neighborhood obesogenic characteristics independently associated with prostate cancer incidence and mortality among Black and White men after controlling for individual prostate cancer risk factors. Multiple neighborhood indices have been developed for each participant, including nSES, the retail food environment index, the restaurant environment index, walkability, and the number of parks, recreation facilities, and businesses. Analyses control for individual-level demographics to evaluate the potential for differential prostate cancer detection. Results may identify neighborhood-level risk factors contributing to race differences in prostate cancer mortality and provide new insights toward reducing obesity and prostate cancer in Black and White people.

# Methods

## Source of data

We used SCCS data for this research. The SCCS is an ongoing cohort study aimed at examining health disparities, including cancer care disparities, among predominantly low-income populations. Participants were recruited into the SCCS between 2002 and 2009, where a total of 84,508 participants aged 40–79 years were enrolled in the cohort. Approximately 85% of the cohort participants were recruited from community health centers, while the remaining 15% were recruited by mail. Details of the SCCS can be found in studies published elsewhere (15, 16). The SCCS received ethical approval from the institutional review boards at Vanderbilt University and Meharry Medical College. All study participants provided written informed consent prior to participation.

**Abbreviations:** aHR, Adjusted hazard ratio; BMI, Body mass index; MEC, Multiethnic cohort; nSES, Neighborhood socioeconomic status; OSM, OpenStreetMap; PSA, Prostate-specific antigen; SCCS, Southern Community Cohort Study; SoDH, Social determinants of health.

In addition, this study was reviewed by the Institutional Review Board of the University of Tennessee Health Science Center for analysis without personal identifiers and was granted a waiver.

## Source of neighborhood-level factors

Neighborhood-level factors were extracted from three primary sources. The first involved sociodemographic data sources from 2010 census data at the block group level. Block groups are statistical divisions smaller than census tracts, contain between 600 and 3,000 people, and are often used for reporting housing and population data [www.census.gov]. Data for estimated median gross rent, education, unemployment rate, median household income, poverty, and house value index were included in this study. The choice of 2010 census data ensured alignment with the SCCS database.

The subsequent data source was the built environment information extracted from OpenStreetMap (OSM). OSM, an open-access, editable global map, encompasses data regarding roads, shopping stores, cafes, and more. Leveraging the overpass application programming interface (API) (https://overpass-turbo.eu/), a robust web tool for querying and retrieving OSM data based on specific attributes, we acquired information on restaurants, retail food establishments, business counts, and recreational facilities in the vicinity of a 1-mile walking catchment of the central point of each block group. ArcGIS Pro 2.5 software was used for spatial data collection.

Lastly, we accessed the walkability index at the block group level from the national walkability index database (https://catalog.data.gov/dataset/walkability-index). This index employs metrics such as street intersection density, proximity to transit stops, and land use diversity to categorize areas into four walkability levels, ranging from minimally walkable to highly walkable.

## Neighborhood obesogenic attributes

The neighborhood socioeconomic environment was a composite measure created by principal component analysis of census block data on housing (median rent and median house value), occupation (proportion with a blue-collar job and proportion older than 16 in a workforce without a job), education (percentage of high school graduates by the year needed to complete high school), employment, and income (median income and percentage of living below the poverty level) (14, 17). The nSES was categorized into quintiles of the distribution, with quintile 1 representing the least economically wealthy neighborhoods and quintile 5 representing the most economically wealthy neighborhoods.

The neighborhood built attributes were the restaurant environment index [the ratio of a fast-food restaurant (e.g., Burger King and McDonald's) to other restaurants (no fast-food and other restaurants): None, quintile 1, 2, 3, or no other restaurants], the retail food environment index [the ratio of the number of convenience stores, liquor stores, and fast-food restaurants to supermarket and farmer's markets (e.g., Kroger, Sprouts, and Publix): None, quintile 1, 2, 3, or no retail food], the number of businesses, the number of parks, and the number of recreational facilities (14, 18). Higher quintiles for the restaurant and retail food environment indices suggest unhealthier neighborhoods regarding the food outlet conditions. Businesses, parks, and recreational facilities were categorized as *none* (no businesses, no parks, or no recreational facilities) and *some* (any businesses, any parks, or any recreational facilities). The walkability index was categorized as least walkable, below average walkable, above average walkable, and most walkable environment (19).

## Individual-level factors

Individual-level factors known to be related to obesity or prostate cancer included in the analyses (collected during the baseline and follow-up time) were age (continuous), race (White or Black individual), currently working (yes, no, or unknown), marital status (married, separated/divorced/widowed, or single), body mass index [<18.5 kg/m$^2$ (underweight), 18.5–24.9 kg/m$^2$ (normal weight), 25–29.9 kg/m$^2$ (overweight), and ≥30 kg/m$^2$ (obese)], smoking status/pack-year (never-smoker, former smoker/<20 pack-years, former smoker/20+ packs-years, former smoker/pack-years unknown, current smoker/<20 pack-years, current smoker/20+ pack-years, and current smoker/pack-years unknown), household income [less than $15,000, at least $15,000 but <$25,000, at least $25,000 but <$50,000, at least $50,000 but < $100,000, $100,000 or more, or unknown (refused/do not know/missing)], total sitting hours (continuous), ever had a history of diabetes milieus (yes, no, or unknown), family history of prostate cancer (parents and siblings) (yes, no, or unknown), and prostate-specific antigen (PSA) utilization (yes, no, or unknown).

## Outcome assessments

The outcome variables in this study were prostate cancer risk and prostate cancer mortality. The SCCS participants were followed from the time of enrollment until the occurrence of cancer diagnosis, death, emigration, or the end of the follow-up period, whichever came first. Incident cancer cases and deaths were identified through the linkage to state cancer registries [ICD-O-3 C61.9, excluding specific histologies (9590–9989, 9050–9055, and 9140+), and considering invasive behavior behavior_icdo3 = 3] and the National Death Registry, as well as from follow-up surveys when confirmed through examination of medical records.

## Analytic sample

Out of the total 84,508 SCCS participants, 34,166 (40.43%) were men and thus eligible for inclusion in our analysis. We excluded 681 participants due to missing BMI information, 337 participants with unknown marital status, 211 participants with missing smoking status and packs per year data, and 1,441 participants with unknown race or those who were neither White nor Black. Since our primary focus was on assessing the risk of prostate cancer development, we also excluded 529 participants who had already been diagnosed with prostate cancer at the time of enrollment in the cohort. Additionally, we excluded 2,611 participants who had missing data for at least one attribute of nSES or could not be linked to individual-level data. Finally, our analysis included a total of 28,356 participants (Figure 1).

## Statistical analysis

Both descriptive and analytical analyses were conducted. The data were described using frequency, percentage, and measure of central locations and dispersions. The median follow-up time with interquartile range (IQR) was calculated. The association between individual-level and neighborhood-level factors was examined using $\chi^2$ tests. The association between neighborhood obesogenic factors and prostate cancer risk was examined using a multivariable Cox proportional hazard ratio and a corresponding 95% confidence interval. Additional race-specific models (White and Black people) were run given the heterogeneity in prostate cancer risks among White and Black people. We also checked the models with and without BMI to determine its impact on the risk of prostate cancer. Additionally, we conducted a stratified analysis based on participants' BMI categories and race groups. However, we excluded participants with a BMI of less than 18.5 kg/m$^2$ from the stratified analysis due to an insufficient sample size. Each model was controlled for individual-level factors including age at enrollment, smoking status and pack-year, marital status, employment status, household income, BMI, and family history of prostate cancer (parents and siblings).

The risk of mortality due to prostate cancer was also estimated using a multivariable Cox proportional hazard ratio among prostate cancer patients. The model was adjusted for individual-level factors and neighborhood obesogenic attributes. The individual-level factors include age at enrollment, smoking status/pack-year, marital status, employment status, BMI, family history of prostate cancer, total sit hours, and PSA utilization. The neighborhood-level factors include neighborhood socioeconomic status, restaurant environment index, retail food environment index, number of parks, number of recreation facilities, and number of businesses.

All statistical analyses were conducted using Stata version 17.0 (Stata Corp LP, College Station, TX, USA) based on two-sided probability, and $p < 0.05$ was considered statistically significant.

## Results

In the analysis, a total of 28,356 participants were included. Among them, 71.2% were Black, while the remaining 28.8% were

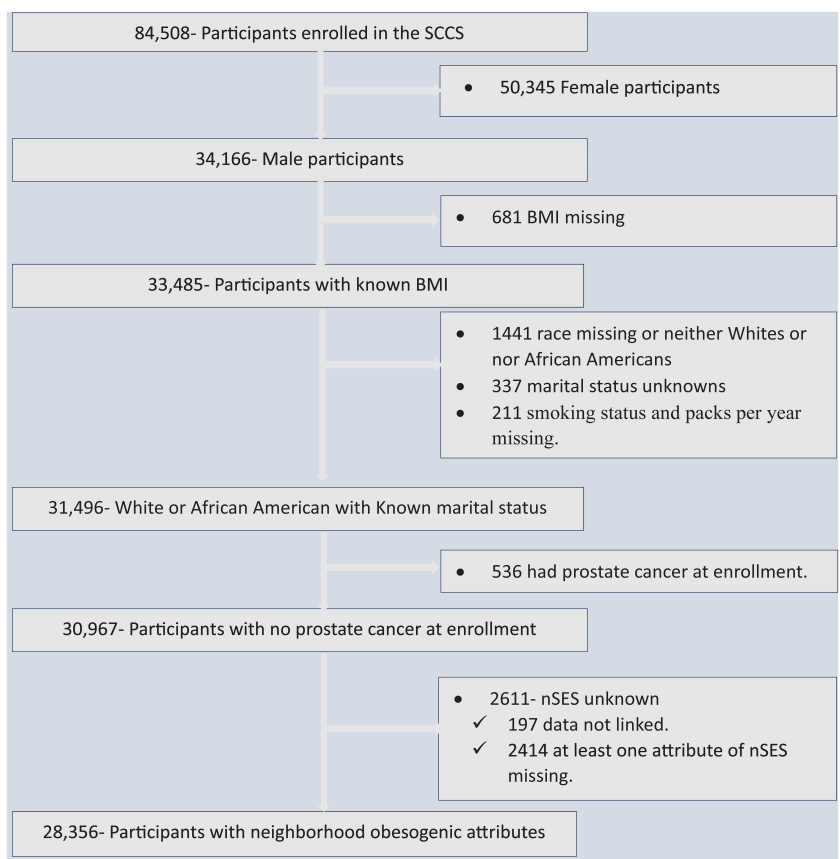

FIGURE 1
Flowchart showing data screening and the inclusion process, the Southern Community Cohort Study, 2002 to 2018. SCCS, the Southern Community Cohort Study; BMI, Body mass index; nSES, neighborhood socioeconomic status.

White. The age of the participants at enrollment ranges from 40 to 79 years with a mean of 51.62 years (53.84 for White and 50.72 for Black individuals). Additionally, 53.36% of White men and 31.72% of Black men were married, while 34.27% of White men and 28.70% of Black men were obese at the time of enrollment (Table 1). The follow-up time ranges from 1 to 177 months with median follow-up time of 133 months (IQR: 103, 152).

Distribution of neighborhood obesogenic attributes varied by racial group. Nearly two-thirds, 64.26% of White people and 58.58% of Black people, lived in a block group without parks. Similarly, 35.36% of White men and 19.51% of Black men lived in the least walkable area. One in four (25.31%) Black men and 10.17% of White men lived in the lowest-nSES quintile (Table 2). Among the participants who lived in the lowest-nSES quintile, 85.78% were Black, while Black individuals only accounted for 56.28% of participants who lived in the highest nSES quintile (Table 3).

The neighborhood obesogenic attributes significantly vary across different BMI categories. More than half of the participants with a BMI of less than 18.5 kg/m$^2$ lived in the lower-quintile (1st and 2nd quintiles) nSES, while only 16.62% lived in the highest

TABLE 1  Participant characteristics across different racial groups, the Southern Community Cohort Study, 2002 to 2018.

| Variables | All<br>n = 28,356 | Black individuals<br>n = 20,186 | White individuals<br>n = 8,170 | p-value |
|---|---|---|---|---|
| | Mean (SD) | Mean (SD) | Mean (SD) | |
| Age at enrollment | 51.62 (8.42) | 50.72 (7.95) | 53.84 (9.11) | <0.001 |
| Total sitting hours | 9.27 (5.17) | 9.34 (5.31) | 9.09 (4.80) | <0.001 |
| Total walk hours | 4.35 (3.70) | 4.52 (3.80) | 3.96 (3.43) | <0.001 |
| | n (%) | n (%) | n (%) | p-value |
| **Family history of prostate cancer** | | | | |
| Yes | 2,588 (9.13) | 1,793 (8.88) | 795 (9.73) | 0.025 |
| No or unknown | 25,768 (90.87) | 18,393 (91.12) | 7,375 (90.27) | |
| **Marital status** | | | | |
| Married | 10,629 (37.48) | 6,357 (31.49) | 4,272 (52.29) | <0.001 |
| Separated/divorced/widowed | 10,248 (36.14) | 7,537 (37.34) | 2,711 (33.18) | |
| Single | 7,479 (26.38) | 6,292 (31.17) | 1,187 (14.53) | |
| **Ever lived in rural or farm** | | | | |
| Yes | 12,868 (45.43) | 8,129 (40.30) | 4,739 (58.10) | <0.001 |
| No | 15,459 (54.57) | 12,042 (59.70) | 3,417 (41.90) | |
| **Household income** | | | | |
| Less than $15,000 | 15,624 (55.10) | 12,076 (59.82) | 3,548 (43.43) | <0.001 |
| At least $15,000 but <$25,000 | 5,700 (20.10) | 4,293 (21.27) | 1,407 (17.22) | |
| At least $25,000 but <$50,000 | 3,859 (13.61) | 2,494 (12.36) | 1,365 (16.71) | |
| At least $50,000 but <$100,000 | 2,114 (7.46) | 903 (4.47) | 1,211 (14.82) | |
| $100,000 or more | 750 (2.64) | 215 (1.07) | 535 (6.55) | |
| Unknown | 309 (1.09) | 205 (1.02) | 104 (1.27) | |
| **Smoking status and packs per year** | | | | |
| Never-smoker | 6,474 (22.83) | 4,438 (21.99) | 2,036 (24.92) | <0.001 |
| Former, less than 20 years | 3,531 (12.45) | 2,461 (12.19) | 1,070 (13.10) | |
| Former, 20 or more years | 2,955 (10.42) | 1,492 (7.39) | 1,463 (17.91) | |
| Former, pack-years unknown | 354 (1.25) | 206 (1.02) | 148 (1.81) | |
| Current, less than 20 years | 7,831 (27.62) | 7,049 (34.92) | 782 (9.57) | |

*(Continued)*

TABLE 1 Continued

| Variables | All<br>n = 28,356 | Black individuals<br>n = 20,186 | White individuals<br>n = 8,170 | p-value |
|---|---|---|---|---|
| | Mean (SD) | Mean (SD) | Mean (SD) | |
| **Smoking status and packs per year** | | | | |
| Current, 20 or more years | 7,086 (24.99) | 4,445 (22.02) | 2,641 (32.33) | |
| Current, pack-years unknown | 125 (0.44) | 95 (0.47) | 30 (0.37) | |
| **BMI** | | | | |
| Underweight | 337 (1.19) | 262 (1.30) | 75 (0.92) | <0.001 |
| Normal | 9,407 (33.17) | 7,101 (35.18) | 2,306 (28.23) | |
| Overweight | 10,074 (35.53) | 7,079 (35.07) | 2,995 (36.66) | |
| Obese | 8,538 (30.11) | 5,744 (28.46) | 2,794 (34.20) | |
| **History of PSA testing** | | | | |
| Yes | 16,897 (59.59) | 11,622 (57.57) | 5,275 (64.57) | <0.001 |
| No or unknown | 11,459 (40.41) | 8,564 (42.43) | 2,895 (35.43) | |

BMI, body mass index; PSA, prostate-specific antigen; SD, standard deviation.

quintile. Fifty-six percent of participants with a normal BMI (18.6–24.9 kg/m$^2$) lived in a neighborhood with no parks, while 63.36% of obese (BMI ≥ 30) participants lived in a similar neighborhood. Moreover, 26.60% of obese participants lived in the least walkable neighborhood, while 9.02% lived in the most walkable neighborhood (Table 4).

The utilization of PSA testing exhibited variation based on race and neighborhood obesogenic attributes. Past PSA testing prevalence was 66% of White and 57% of Black participants. Similarly, within the White men, 35.83% of those residing in the wealthiest quintile of nSES had undergone PSA testing, while only 17.33% of Black men had done so. Furthermore, 6.39% of White people residing in the lowest quintile of the restaurant environment index had undergone PSA testing, compared to 11.37% of their Black counterparts (Table 5).

A total of 1,524(5.37%) participants were diagnosed with prostate cancer, including 313 White men (3.83%) and 1211 Black men (6.00%). After adjusting for individual-level prostate cancer risk factors as well as neighborhood-level attributes, none of the neighborhood's obesogenic factors demonstrated a significant association with prostate cancer risk, except for the lowest quintile of nSES and the retail food environment index. Participants residing in neighborhoods within the lowest quintile of nSES exhibited a 32% higher risk of prostate cancer compared to those residing in the wealthiest quintile of nSES (aHR = 1.32, 95% CI 1.12–1.57, p = 0.001). Furthermore, the retail food environment index showed an association with an elevated risk of prostate cancer (aHR = 1.45, 95% CI 1.04–2.03, p = 0.029, for quintile 1 vs. None) among all participants in general and Black individuals in particular (aHR = 1.53, 95% CI 1.07–2.20, p = 0.021, for quintile 1 vs. None) (Table 6).

After further stratifying based on race and BMI, participants with a normal BMI living in neighborhoods within the lowest quintile of nSES had a 46% higher risk of prostate cancer compared to those in the wealthiest quintile of nSES (aHR = 1.46, 95% CI 1.07–1.98, p = 0.016). This increased risk was primarily observed in Black men (aHR = 1.46, 95% CI 1.03–2.09, p = 0.036), while no statistically significant association between nSES and the risk of prostate cancer was observed among White people with a normal weight (aHR = 0.66, 95% CI 0.25–1.80, p = 0.420). In contrast, overweight participants living in quintile 4 of nSES had a 40% higher risk of prostate cancer compared to those in the wealthiest quintile of nSES (aHR = 1.40, 95% CI 1.09–1.80, p = 0.016). However, no statistically significant association was observed among White and Black individuals in this regard. Furthermore, the restaurant environment index showed a protective effect on the risk of prostate cancer among overweight Black men (aHR = 0.63, 95% CI 0.39–0.99, p = 0.048, for quintile 1 vs. None), while it was associated with an elevated risk among overweight White individuals (aHR = 3.37, 95% CI 1.04–10.94, p = 0.043, for quintile 1 vs. None). On the other hand, the retail food environment index showed an association with an elevated risk of prostate cancer among overweight Black individuals (aHR = 2.27, 95% CI 1.22–4.23, p = 0.010, for quintile 2 vs. None). Obese Black individuals residing in neighborhoods with no recreational facilities had a 42% higher risk of prostate cancer (aHR = 1.42, 95% CI 1.04–1.94, p = 0.026) compared to similar participants residing in neighborhoods with recreational facilities (Table 7).

Similarly, participants residing in neighborhoods within the lower quintiles (quintile 1 and quintile 2) of nSES had an increased risk of prostate cancer-related mortality in a dose-response manner. Participants residing in neighborhoods within quintile 1 of nSES had a 3.45 times higher risk of mortality due to prostate cancer (aHR = 3.45, 95% CI 1.54–7.64, p = 0.003), while those who reside in quintile 2 had a 2.28 times higher risk of prostate cancer-related deaths compared to those residing in the wealthiest quintile of SES (aHR = 2.28, 95% CI 1.01–5.12, $p$ = 0.043). Participants who ever performed PSA testing had a 54% lower risk of death due to prostate cancer compared to participants who did not have a prior PSA test (aHR = 0.46, 95% CI 0.28–0.75, $p$ = 0.002). Moreover, the restaurant environment index showed an association with an elevated risk of prostate cancer mortality (aHR = 5.12, 95% CI 1.57–16.67, $p$ = 0.007, for quintile 1 vs. None). Participants residing in the least walkable environment had a 3.45 higher risk of death due to prostate cancer compared to participants residing in the most walkable environment (aHR = 3.45, 95% CI 1.22–9.78, $p$ = 0.020) (Table 8).

TABLE 2 Neighborhood obesogenic factors across different races, the Southern Community Cohort Study, 2002 to 2018.

| Variables | All, n (%) n = 28,356 | Black individuals, n (%) n = 20,186 | White individuals, n (%) n = 8,170 | p-value |
|---|---|---|---|---|
| **nSES** | | | | |
| Quantile 1 | 5,964 (21.03) | 5,116 (25.34) | 848 (10.38) | <0.001 |
| Quantile 2 | 5,621 (19.82) | 4,525 (22.42) | 1,096 (13.41) | |
| Quantile 3 | 5,298 (18.68) | 3,740 (18.53) | 1,558 (19.07) | |
| Quantile 4 | 5,393 (19.02) | 3,383 (16.76) | 2,010 (24.60) | |
| Quantile 5 | 6,080 (21.44) | 3,422 (16.95) | 2,658 (32.53) | |
| **Restaurant environment index** | | | | |
| None[a] | 19,680 (69.40) | 13,915 (68.93) | 5,765 (70.56) | <0.001 |
| Quantile 1 | 3,276 (11.55) | 2,611 (12.93) | 665 (8.14) | |
| Quantile 2 | 863 (3.04) | 578 (2.86) | 285 (3.49) | |
| Quantile 3 | 1,473 (5.19) | 1,071 (5.31) | 402 (4.92) | |
| No restaurant | 3,064 (10.81) | 2,011 (9.96) | 1,053 (12.89) | |
| **Retail food environment index** | | | | |
| None[b] | 14,212 (50.12) | 9,172 (45.44) | 5,040 (61.69) | <0.001 |
| Quantile 1 | 981 (3.46) | 728 (3.61) | 253 (3.10) | |
| Quantile 2 | 850 (3.00) | 657 (3.25) | 193 (2.36) | |
| Quantile 3 | 1,208 (4.26) | 944 (4.68) | 264 (3.23) | |
| No retail food | 11,105 (39.16) | 8,685 (43.02) | 2,420 (29.62) | |
| **Number of parks** | | | | |
| None | 16,936 (59.73) | 11,836 (58.63) | 5,100 (62.42) | <0.001 |
| Some | 11,420 (40.27) | 8,350 (41.37) | 3,070 (37.58) | |
| **Number of recreation facilities** | | | | |
| None | 17,990 (63.44) | 12,615 (62.49) | 5,375 (65.79) | <0.001 |
| Some | 10,366 (36.56) | 7,571 (37.51) | 2,795 (34.21) | |
| **Number of businesses** | | | | |
| None | 23,888 (84.24) | 17,150 (84.96) | 6,738 (82.47) | <0.001 |
| Some | 4,468 (15.76) | 3,036 (15.04) | 1,432 (17.53) | |
| **Walkability index** | | | | |
| Least walkable | 6,268 (22.10) | 3,625 (17.96) | 2,643 (32.35) | <0.001 |
| Below average | 10,210 (36.01) | 7,077 (35.06) | 3,133 (38.35) | |
| Above average | 8,364 (29.50) | 6,675 (33.07) | 1,689 (20.67) | |
| Most walkable | 3,514 (12.39) | 2,809 (13.92) | 705 (8.63) | |

[a]No fast-food restaurant and other restaurants; [b]No fast food restaurant and retail food. nSES, neighborhood socioeconomic status.

TABLE 3 Distribution of individual-level factors across different nSES quintiles, the Southern Community Cohort Study, 2002 to 2018.

| Variable | nSES | | | | | p-value |
|---|---|---|---|---|---|---|
| | Quintile 1, % n = 5,964 | Quintile 2, % n = 5,621 | Quintile 3, % n = 5,298 | Quintile 4, % n = 5,393 | Quintile 5, % n = 5,393 | |
| Race | | | | | | |
| White | 14.22 | 19.50 | 29.41 | 37.27 | 43.72 | <0.001 |
| Black | 85.78 | 80.50 | 70.59 | 62.73 | 56.28 | |
| Age at enrollment | | | | | | |
| 40–49 | 50.57 | 48.02 | 46.24 | 45.78 | 44.64 | <0.001 |
| 50–59 | 33.45 | 35.12 | 35.50 | 35.36 | 35.38 | |
| 60–69 | 12.76 | 13.08 | 14.48 | 15.15 | 16.15 | |
| 70–79 | 3.22 | 3.79 | 3.78 | 3.71 | 3.83 | |
| Marital Status | | | | | | |
| Married | 32.60 | 34.64 | 37.96 | 40.76 | 41.60 | <0.001 |
| Separated/divorced/widowed | 37.66 | 37.38 | 35.52 | 35.99 | 34.18 | |
| Single | 29.75 | 27.98 | 26.52 | 23.25 | 24.23 | |
| Household income | | | | | | |
| Less than $15,000 | 66.83 | 60.45 | 54.21 | 50.12 | 43.83 | <0.001 |
| At least $15,000 but <$25,000 | 19.79 | 21.21 | 21.76 | 20.12 | 17.93 | |
| At least $25,000 but <$50,000 | 9.56 | 12.15 | 15.67 | 16.02 | 15.00 | |
| At least $50,000 but <$100,000 | 2.46 | 4.54 | 6.21 | 10.11 | 13.78 | |
| $100,000 or more | 0.39 | 0.75 | 1.13 | 2.41 | 8.14 | |
| Unknown | 0.97 | 0.91 | 1.02 | 1.22 | 1.32 | |
| Smoking status and pack-year | | | | | | |
| Never-smoker | 21.71 | 21.08 | 21.86 | 23.66 | 25.66 | <0.001 |
| Former, less than 20 years | 11.08 | 12.17 | 12.78 | 12.94 | 13.34 | |
| Former, 20 or more years | 8.99 | 9.66 | 11.00 | 11.37 | 11.18 | |
| Former, pack-years unknown | 1.06 | 1.14 | 1.23 | 1.37 | 1.45 | |
| Current, less than 20 years | 31.87 | 30.71 | 26.37 | 24.11 | 24.79 | |
| Current, 20 or more years | 24.82 | 24.82 | 26.20 | 26.22 | 23.17 | |
| Current, pack-years unknown | 0.47 | 0.43 | 0.57 | 0.33 | 0.41 | |
| BMI | | | | | | |
| Underweight | 1.56 | 1.48 | 0.98 | 0.98 | 0.92 | <0.001 |
| Normal | 34.37 | 35.24 | 31.77 | 30.58 | 33.62 | |
| Overweight | 34.37 | 33.50 | 35.67 | 36.77 | 37.30 | |
| Obese | 29.69 | 29.78 | 31.58 | 31.67 | 28.16 | |
| PSA | | | | | | |
| Yes | 55.28 | 56.75 | 60.42 | 61.71 | 63.83 | <0.001 |
| No or unknown | 44.72 | 43.25 | 39.58 | 38.29 | 36.17 | |
| Family history of prostate cancer | | | | | | |
| Yes | 8.27 | 8.41 | 9.25 | 9.55 | 10.15 | 0.001 |
| No or Unknown | 91.73 | 91.59 | 90.75 | 90.45 | 89.85 | |

(Continued)

TABLE 3 Continued

| Variable | nSES | | | | | p-value |
|---|---|---|---|---|---|---|
| | Quintile 1, % n = 5,964 | Quintile 2, % n = 5,621 | Quintile 3, % n = 5,298 | Quintile 4, % n = 5,393 | Quintile 5, % n = 5,393 | |
| Prostate cancer | | | | | | |
| Yes | 5.89 | 4.84 | 5.23 | 5.66 | 5.25 | 0.112 |
| No or unknown | 94.11 | 95.16 | 94.77 | 94.34 | 94.75 | |

BMI, body mass index; PSA, prostate-specific antigen.

# Discussion

Our investigation of neighborhood obesogenic factors with prostate cancer included over 28,000 White and Black men living in the southeastern United States. We observed that low nSES was significantly associated with an increase in overall prostate cancer risk and mortality in Black men in the southeastern United States. However, we detected variation based on participants' BMI categories. Among Black men with normal weight, low (quintile 1) nSES was significantly associated with an increase in prostate cancer risk. Among overweight Black people, higher (quintile 4) nSES was significantly associated with an increase in prostate cancer risk. Obese Black people residing in neighborhoods with no recreational facilities also had a higher risk of prostate cancer. Furthermore, PSA testing was more common among White men with a high nSES compared to Black men with a high nSES. We also found that prostate cancer risk was associated with lower levels of the retail food environment, which was specific to Black men. The low restaurant environment index showed a protective effect on the risk of prostate cancer among overweight Black men, while it was associated with an elevated risk among overweight White people. Walkability had no statistically significant association with risk of prostate cancer development but with risk mortality from prostate cancer.

Increasing socioeconomic status has been associated with increased PSA testing, potentially inducing a detection bias leading to the appearance that increasing SES increases prostate cancer risk (20). Indeed, we also observed increased PSA testing associated with increased nSES and controlled for past PSA testing practices in our analyses of the neighborhood-built environment. In contrast to the expectation that any relationship was driven solely by a selective detection, we found increased nSES to be significantly associated with a lower prostate cancer risk. This relationship was specific to Black men while increasing nSES was associated with a non-significant increase in prostate cancer risk among White men. However, a contradictory finding reported by a previous study indicated that higher nSES was associated with higher prostate cancer risk (14), with the association solely observed among foreign-born Latino men with non-aggressive disease. Although SCCS recruitment included an overrepresentation of lower-income participants overall, Black individuals in this study had a lower nSES and a lower PSA testing level than White individuals.

Why the relationship between nSES and prostate cancer was driven largely by an increased risk among Black men is not clear. However, evidence suggests that social inequities, such as structural racism and mistrust, including a history of segregation and mistreatment within the healthcare system, could impact Black men's access to screening and treatment for prostate cancer, as well as their participation in clinical research, ultimately influencing prostate cancer outcomes and survival (21–23). Further evidence shows that structural racism is one of the major barriers to health equity (24, 25). In our analysis, race designations are evaluated as social rather than biological constructs. The neighborhood-built environment is often segregated by race as much as by socioeconomic status. Area resources affect all aspects of life, including the location of stores offering unhealthy food. The availability of healthy food choices may serve as opportunities to communicate with stakeholders and as intervention targets for future policies of community health outcomes. Poorer areas may also have fewer public spaces, which may reduce opportunities to exercise. New data collection in future studies should also investigate neighborhood systems that might contain reinforcing or counterbalancing components that could compensate for deprivation or any alternation in the system.

Indeed, our analysis also found that less availability of healthy foods increased prostate risk among Black men. While associations were not always statistically significant, our findings reveal that individuals' health outcomes are influenced differently by their neighborhood experiences, depending on their racial backgrounds. For example, as the retail food environment index decreases, the prostate cancer risk among Black men increases. However, as the restaurant environment index increases, the prostate cancer risk among Black people increases but decreased among White people. Similarly, our finding showed that the restaurant environment index was associated with prostate cancer mortality. A recent study revealed that counties exhibiting the highest food swamp scores (determined by the ratio of fast-food restaurants and convenience stores to grocery stores and farmer's markets) and food desert conditions (quantified by the proportion of each county's population characterized by both low income and limited access to grocery stores (26)) experienced higher odds of obesity-related cancer mortality compared to counties with lower food swamp and food desert scores (27). Considering the existence of interactions between race, SDOH, and prostate cancer risk and survival, a recent systematic review and meta-analysis underscored the importance of incorporating SDOH, including neighborhood-level attributes, into research on racial disparities in prostate cancer (28).

Walkability showed no statistically significant association with the risk of developing prostate cancer. However, we noted a non-statistically significant trend of increased prostate cancer risk among overweight and normal-weight Black men as the walkability index decreased. Conversely, we observed a non-statistically significant protective effect on the risk of prostate cancer among White men with normal weight as the walkability index decreased, and an increase in prostate cancer risk among obese White men as the walkability index increased. Participants residing in the least walkable nSES, neighborhood socioeconomic status; [a]No fast-food restaurant and other restaurants; [b]No fast-food restaurant and retail food.

TABLE 4 Neighborhood obesogenic factors across different BMI groups, the Southern Community Cohort Study, 2002 to 2018.

| Variable | Underweight, %, n = 337 | Normal weight, %, n = 9,407 | Overweight, %, n = 10,074 | Obese, %, n = 8,538 | p-value |
|---|---|---|---|---|---|
| **nSES** | | | | | |
| Quintile 1 | 27.60 | 21.79 | 20.35 | 20.74 | <0.001 |
| Quintile 2 | 24.63 | 21.06 | 18.69 | 19.61 | |
| Quintile 3 | 15.43 | 17.89 | 18.76 | 19.59 | |
| Quintile 4 | 15.73 | 17.53 | 19.68 | 20.00 | |
| Quintile 5 | 16.62 | 21.73 | 22.51 | 20.05 | |
| **Restaurant environment index** | | | | | |
| None[a] | 65.28 | 65.95 | 69.75 | 72.96 | <0.001 |
| Quintile 1 | 16.91 | 14.16 | 11.18 | 8.91 | |
| Quintile 2 | 4.45 | 3.22 | 3.30 | 2.49 | |
| Quintile 3 | 4.15 | 5.86 | 5.29 | 4.39 | |
| No restaurant | 9.20 | 10.81 | 10.48 | 11.24 | |
| **Retail food environment index** | | | | | |
| None[b] | 41.84 | 42.69 | 51.23 | 57.32 | <0.001 |
| Quintile 1 | 4.45 | 4.03 | 3.25 | 3.05 | |
| Quintile 2 | 3.26 | 3.32 | 3.12 | 2.49 | |
| Quintile 3 | 4.75 | 5.47 | 4.21 | 2.96 | |
| No retail food | 45.70 | 44.49 | 38.20 | 34.18 | |
| **Number of parks** | | | | | |
| Some | 45.40 | 43.84 | 39.86 | 36.64 | <0.001 |
| None | 54.60 | 56.16 | 60.14 | 63.36 | |
| **Number of recreation facilities** | | | | | |
| Some | 37.09 | 40.87 | 36.33 | 32.04 | <0.001 |
| None | 62.91 | 59.13 | 63.67 | 67.96 | |
| **Number of businesses** | | | | | |
| None | 85.76 | 83.58 | 84.09 | 85.10 | <0.001 |
| Some | 14.24 | 16.42 | 15.91 | 14.90 | |
| **Walkability index** | | | | | |
| Least walkable | 18.99 | 17.63 | 22.58 | 26.60 | <0.001 |
| Below average | 35.61 | 33.10 | 36.64 | 38.48 | |
| Above average | 29.67 | 33.55 | 28.75 | 25.91 | |
| Most walkable | 15.73 | 15.72 | 12.03 | 9.02 | |

[a]No fast-food restaurant and other restaurants; [b]No fast-food restaurant and retail food; nSES, neighborhood socioeconomic status.

TABLE 5 Neighborhood-level factors and PSA across different races, the Southern Community Cohort Study, 2002 to 2018.

| Variable | All | | Black individuals | | White individuals | |
|---|---|---|---|---|---|---|
| | Yes, % $n = 16{,}897$ | No or unknown, % $n = 11{,}459$ | PSA | | | |
| | | | Yes, % $n = 11{,}622$ | No or unknown, % $n = 8{,}564$ | Yes, % $n = 5{,}275$ | No or unknown, % $n = 2{,}895$ |
| **nSES** | | | | | | |
| Quintile 1 | 19.51 | 23.27 | 24.24 | 26.84 | 9.10 | 12.71 |
| Quintile 2 | 18.88 | 21.21 | 21.97 | 23.03 | 12.08 | 15.85 |
| Quintile 3 | 18.94 | 18.30 | 18.99 | 17.90 | 18.84 | 19.48 |
| Quintile 4 | 19.70 | 18.02 | 17.54 | 15.71 | 24.45 | 24.87 |
| Quintile 5 | 22.97 | 19.19 | 17.27 | 16.52 | 35.53 | 27.08 |
| **Restaurant environment index** | | | | | | |
| None[a] | 71.00 | 67.05 | 70.31 | 67.07 | 72.53 | 66.98 |
| Quintile 1 | 10.40 | 13.26 | 11.95 | 14.27 | 6.98 | 10.26 |
| Quintile 2 | 2.85 | 3.32 | 2.66 | 3.14 | 3.28 | 3.87 |
| Quintile 3 | 4.91 | 5.61 | 5.09 | 5.59 | 4.51 | 5.66 |
| No restaurant | 10.84 | 10.76 | 9.99 | 9.93 | 12.70 | 13.23 |
| **Retail food environment index** | | | | | | |
| None[b] | 53.68 | 44.87 | 48.62 | 41.11 | 64.82 | 55.99 |
| Quintile 1 | 2.83 | 4.39 | 3.09 | 4.31 | 2.26 | 4.63 |
| Quintile 2 | 2.65 | 3.51 | 2.90 | 3.74 | 2.10 | 2.83 |
| Quintile 3 | 3.79 | 4.95 | 4.29 | 5.20 | 2.69 | 4.21 |
| No retail food | 37.05 | 42.28 | 41.09 | 45.64 | 28.13 | 32.33 |
| **Number of parks** | | | | | | |
| None | 61.77 | 56.71 | 60.27 | 56.42 | 65.10 | 57.55 |
| Some | 38.23 | 43.29 | 39.73 | 43.58 | 34.90 | 42.45 |
| **Number of recreation facilities** | | | | | | |
| None | 66.41 | 59.06 | 65.28 | 58.71 | 68.91 | 60.10 |
| Some | 33.59 | 40.94 | 34.72 | 41.29 | 31.09 | 39.90 |
| **Number of businesses** | | | | | | |
| None | 84.83 | 83.38 | 85.40 | 84.36 | 83.56 | 80.48 |
| Some | 15.17 | 16.62 | 14.60 | 15.64 | 16.44 | 19.52 |
| **Walkability index** | | | | | | |
| Least walkable | 25.04 | 17.78 | 20.47 | 14.55 | 35.11 | 27.32 |
| Below average | 38.18 | 32.80 | 36.95 | 32.50 | 40.89 | 33.71 |
| Above average | 26.48 | 33.95 | 30.49 | 36.56 | 17.63 | 26.22 |
| Most walkable | 10.30 | 15.47 | 12.09 | 16.39 | 6.37 | 12.75 |

neighborhoods had a higher likelihood of mortality from prostate cancer compared to those living in the most walkable neighborhoods.

Lower nSES was also associated with increased prostate cancer mortality. A similar finding has been reported in a previous study (29). Additionally, residing in a neighborhood characterized by reduced walkability is associated with an elevated risk of prostate cancer-related mortality. A previous study reported that greater neighborhood walkability was associated with lower BMI among African American cancer survivors (30). Walking has widespread impacts on various aspects, such as metabolism, insulin sensitivity, reduced body fat, enhanced mental wellbeing, decreased stress, and emotional health. Limited walking may contribute to a poor quality of life and unfavorable outcomes for individuals diagnosed with prostate cancer. Prostate cancer mortality was also significantly associated with lower PSA testing. Previous studies also reported improved cancer-specific survival in the PSA era compared with the

TABLE 6 Neighborhood obesogenic factors and risk of prostate cancer, the Southern Community Cohort Study, 2002 to 2018.

| Variables | All | | Black individuals | | White individuals | |
|---|---|---|---|---|---|---|
| | Cases n = 1,524 | aHR (95% CI) | Cases n = 1,211 | aHR (95% CI) | Cases n = 313 | aHR (95% CI) |
| **nSES** | | | | | | |
| Quintile 5 (high) | 319 | Reference | 195 | Reference | 124 | Reference |
| Quintile 4 | 305 | 1.18 [0.97–1.34] | 226 | 1.17 [0.96–1.43] | 79 | 1.06 [0.79–1.43] |
| Quintile 3 | 277 | 1.14 [0.96–1.36] | 224 | 1.07 [0.88–1.32] | 53 | 1.02 [0.73–1.44] |
| Quintile 2 | 272 | 1.06 [0.89–1.26] | 235 | 0.94 [0.77–1.16] | 37 | 0.97 [0.73–1.45] |
| Quintile 1 (low) | 351 | 1.32 [1.12–1.57]* | 331 | 1.16 [0.96–1.41] | 20 | 0.81 [0.49–1.33] |
| **Restaurant environment index** | | | | | | |
| None[a] | 1,078 | Reference | 850 | Reference | 228 | Reference |
| Quintile 1 | 169 | 1.00 [0.78–1.29] | 140 | 0.92 [0.70–1.20] | 29 | 2.06 [0.98–4.33] |
| Quintile 2 | 35 | 0.88 [0.59–1.30] | 25 | 0.95 [0.77–1.16] | 10 | 1.32 [0.53–3.26] |
| Quintile 3 | 65 | 0.90 [0.64–1.26] | 55 | 0.98 [0.68–1.43] | 10 | 0.88 [0.36–2.20] |
| No restaurant | 177 | 0.97 [0.76–1.24] | 141 | 1.06 [0.82–1.37] | 36 | 1.38 [0.68–2.79] |
| **Retail food environment index** | | | | | | |
| None[b] | 787 | Reference | 578 | Reference | 209 | Reference |
| Quintile 1 | 49 | 1.45 [1.04–2.03]* | 42 | 1.53 [1.07–2.20]* | 7 | 0.66 [0.27–1.66] |
| Quintile 2 | 39 | 1.22 [0.84–1.78] | 35 | 1.32 [0.88–1.98] | 4 | 0.39 [0.13–1.23] |
| Quintile 3 | 57 | 1.18 [0.83–1.68] | 49 | 1.12 [0.76–1.66] | 8 | 0.65 [0.26–1.62] |
| No retail food | 592 | 1.07 [0.92–1.24] | 507 | 1.05 [0.89–1.23] | 85 | 0.68 [0.42–1.11] |
| **Number of parks** | | | | | | |
| Some | 581 | Reference | 477 | Reference | 104 | Reference |
| None | 943 | 0.92 [0.76–1.12] | 734 | 0.89 [0.72–1.10] | 209 | 1.16 [0.70–1.97] |
| **Number of recreation facilities** | | | | | | |
| Some | 492 | Reference | 396 | Reference | 96 | Reference |
| None | 1,032 | 1.09 [0.94–1.27] | 815 | 1.10 [0.94–1.30] | 217 | 0.94 [0.64–1.37] |
| **Number of businesses** | | | | | | |
| None | 1,312 | Reference | 1,052 | Reference | 260 | Reference |
| Some | 212 | 0.98 [0.83–1.14] | 159 | 0.94 [0.79–1.13] | 53 | 1.18 [0.83–1.70] |
| **Walkability index** | | | | | | |
| Least walkable | 368 | 1.08 [0.85–1.36] | 253 | 1.23 [0.95–1.60] | 114 | 0.82 [0.46–1.45] |
| Below average | 568 | 1.10 [0.89–1.35] | 454 | 1.22 [0.97–1.54] | 114 | 0.75 [0.44–1.27] |
| Above average | 427 | 1.09 [0.90–1.33] | 364 | 1.09 [0.88–1.36] | 63 | 1.04 [0.62–1.74] |
| Most walkable | 161 | Reference | 139 | Reference | 22 | Reference |

Controlled for age at enrollment (continuous), smoking status and pack-year, marital status, employment status, household income, body mass index at enrollment, family history of prostate cancer, neighborhood socioeconomic status, restaurant environment index, retail food environment index, number of parks, number of recreation facilities, number of businesses, and walkability index. [a]No fast-food restaurant and other restaurants; [b]No fast-food restaurant and retail food; nSES, neighborhood socioeconomic status; * statistically significant (p-value < 0.05); p-interactions for the associations race and BMI <0.001.

TABLE 7 Neighborhood obesogenic factors and risk of prostate cancer stratified by BMI categories and race, the Southern Community Cohort Study, 2002 to 2018.

| Variables | All | Black individuals | White individuals |
|---|---|---|---|
| | aHR (95% CI) | aHR (95% CI) | aHR (95% CI) |
| Normal weight (BMI: 18.5–24.9 kg/m$^2$) | | | |
| nSES | | | |
| Quintile 5 (high) | Reference | Reference | Reference |
| Quintile 4 | 1.07 [0.98–1.46] | 1.25 [0.82–1.81] | 0.65 [0.35–1.22] |
| Quintile 3 | 1.26 [0.92–1.72] | 1.19 [0.82–1.73] | 1.13 [0.60–2.13] |
| Quintile 2 | 1.07 [0.78–1.46] | 1.01 [0.70–1.46] | 0.85 [0.40–1.78] |
| Quintile 1 (low) | 1.46 [1.07–1.98]* | 1.46 [1.03–2.09]* | 0.66 [0.25–1.80] |
| Restaurant environment index | | | |
| None[a] | Reference | Reference | Reference |
| Quintile 1 | 1.21 [0.78–1.86] | 1.26 [0.80–1.99] | 0.56 [0.13–2.50] |
| Quintile 2 | 0.83 [0.39–1.76] | 0.87 [0.35–2.14] | 0.85 [0.17–4.29] |
| Quintile 3 | 1.34 [0.77–2.34] | 1.67 [0.91–3.06] | 0.70 [0.16–3.12] |
| No restaurant | 1.04 [0.66–1.62] | 1.13 [0.70–1.81] | 0.96 [0.28–3.26] |
| Retail food environment index | | | |
| None[b] | Reference | Reference | Reference |
| Quintile 1 | 1.48 [0.86–2.55] | 1.54 [0.87–2.74] | 0.62 [0.07–5.30] |
| Quintile 2 | 1.14 [0.61–2.12] | 1.01 [0.52–1.99] | 1.42 [0.26–7.82] |
| Quintile 3 | 0.87 [0.49–1.53] | 0.77 [0.41–1.45] | 1.02 [0.24–4.37] |
| No retail food | 1.03 [0.79–1.24] | 1.04 [0.77–1.39] | 0.95 [0.43–2.10] |
| Number of parks | | | |
| Some | Reference | Reference | Reference |
| None | 0.97 [0.67–1.40] | 0.91 [0.61–1.36] | 1.31 [0.47–3.66] |
| Number of recreation facilities | | | |
| Some | Reference | Reference | Reference |
| None | 1.04 [0.79–1.36] | 0.99 [0.74–1.33] | 1.08 [0.52–2.26] |
| Number of businesses | | | |
| None | Reference | Reference | Reference |
| Some | 1.13 [0.86–1.49] | 1.16 [0.86–1.58] | 1.36 [0.67–2.78] |
| Walkability index | | | |
| Least walkable | 1.06 [0.70–1.60] | 1.37 [0.87–2.16] | 0.40 [0.14–1.16] |
| Below average | 1.04 [0.94–1.47] | 1.23 [0.84–1.79] | 0.44 [0.17–1.14] |
| Above average | 0.95 [0.69–1.30] | 1.04 [0.73–1.46] | 0.49 [0.20–1.21] |
| Most walkable | Reference | Reference | Reference |
| Overweight (BMI: 25–29.9 kg/m$^2$) | | | |
| nSES | | | |
| Quintile 5 (high) | Reference | Reference | Reference |
| Quintile 4 | 1.40 [1.09–1.80]* | 1.32 [0.96–1.81] | 1.44 [0.93–2.25] |

*(Continued)*

TABLE 7 Continued

| Variables | All | Black individuals | White individuals |
|---|---|---|---|
|  | aHR (95% CI) | aHR (95% CI) | aHR (95% CI) |
| **nSES** | | | |
| Quintile 3 | 0.94 [0.70–1.25] | 0.93 [0.66–1.30] | 0.72 [0.39–1.31] |
| Quintile 2 | 1.03 [0.77–1.37] | 0.91 [0.65–1.28] | 1.08 [0.56–2.07] |
| Quintile 1 (low) | 1.30 [0.99–1.72] | 1.14 [0.83–1.57] | 0.79 [0.31–1.92] |
| **Restaurant environment Index** | | | |
| None[a] | Reference | Reference | Reference |
| Quintile 1 | 0.79 [0.53–1.19] | 0.63 [0.39–0.99] | 3.37 [1.04–10.94]* |
| Quintile 2 | 0.75 [0.41–1.37] | 0.98 [0.51–1.90] | 0.83 [0.17–4.12] |
| Quintile 3 | 0.55 [0.31–0.98] | 0.57 [0.30–1.09] | 0.73 [0.16–3.27] |
| No restaurant | 0.82 [0.55–1.20] | 0.99 [0.65–1.51] | 1.07 [0.32–3.55] |
| **Retail food environment Index** | | | |
| None[b] | Reference | Reference | >Reference |
| Quintile 1 | 1.52 [0.86–2.69] | 1.69 [0.90–3.18] | 0.47 [0.10–2.19] |
| Quintile 2 | 1.63 [0.90–2.95] | 2.27 [1.22–4.23] | |
| Quintile 3 | 1.53 [0.85–2.74] | 1.39 [0.72–2.71] | 1.09 [0.25–4.72] |
| No retail food | 1.16 [0.91–1.47] | 1.11 [0.86–1.43] | 0.59 [0.23–1.49] |
| **Number of parks** | | | |
| Some | Reference | Reference | Reference |
| None | 0.85 [0.62–1.16] | 0.90 [0.64–1.28] | 0.61 [0.29–1.30] |
| **Number of recreation facilities** | | | |
| Some | Reference | Reference | Reference |
| None | 1.01 [0.78–1.29] | 1.00 [0.76–1.32] | 1.19 [0.65–2.21] |
| **Number of businesses** | | | |
| None | Reference | Reference | Reference |
| Some | 0.96 [0.74–1.24] | 0.81 [0.60–1.11] | 1.41 [0.81–2.46] |
| **Walkability index** | | | |
| Least walkable | 1.03 [0.70–1.52] | 1.04 [0.67–1.62] | 1.17 [0.46–2.98] |
| Below average | 0.95 [0.67–1.34] | 1.02 [0.70–1.50] | 0.73 [0.31–1.73] |
| Above average | 1.01 [0.73–1.41] | 0.99 [0.69–1.44] | 0.98 [0.42–2.29] |
| Most walkable | Reference | Reference | Reference |
| **Obese (BMI $\geq$ 30 kg/m$^2$)** | | | |
| **nSES** | | | |
| Quintile 5 (high) | Reference | Reference | Reference |
| Quintile 4 | 1.04 [0.77–1.41] | 0.95 [0.66–1.37] | 1.05 [0.60–1.86] |
| Quintile 3 | 1.23 [0.92–1.66] | 1.05 [0.74–1.50] | 1.33 [0.75–2.37] |
| Quintile 2 | 1.07 [0.79–1.46] | 0.90 [0.63–1.29] | 1.10 [0.55–2.20] |
| Quintile 1 (low) | 1.21 [0.90–1.64] | 0.94 [0.67–1.34] | 1.02 [0.46–2.23] |

(Continued)

TABLE 7 Continued

| Variables | All | Black individuals | White individuals |
|---|---|---|---|
| | aHR (95% CI) | aHR (95% CI) | aHR (95% CI) |
| **Restaurant environment index** | | | |
| None[a] | Reference | Reference | Reference |
| Quintile 1 | 1.01 [0.61–1.66] | 0.89 [0.52–1.52] | 2.48 [0.53–11.50] |
| Quintile 2 | 1.19 [0.57–2.50] | 0.90 [0.36–2.26] | 5.27 [0.98–28.31] |
| Quintile 3 | 0.74 [0.36–1.54] | 0.71 [0.32–1.58] | 0.90 [0.08–9.54] |
| No restaurant | 1.12 [0.72–1.74] | 1.09 [0.68–1.74] | 3.37 [0.83–13.73] |
| **Retail food environment index** | | | |
| None[b] | Reference | Reference | Reference |
| Quintile 1 | 1.40 [0.73–2.68] | 1.39 [0.67–2.86] | 0.98 [0.22–4.46] |
| Quintile 2 | 0.97 [0.43–2.16] | 0.94 [0.37–2.39] | 0.56 [0.09–3.26] |
| Quintile 3 | 1.16 [0.54–2.47] | 1.54 [0.68–3.50] | |
| No retail food | 1.03 [0.78–1.35] | 1.02 [0.76–1.37] | 0.63 [0.26–1.48] |
| **Number of parks** | | | |
| Some | Reference | Reference | Reference |
| None | 0.95 [0.67–1.36] | 0.80 [0.55–1.17] | 2.84 [0.98–8.25] |
| **Number of recreation facilities** | | | |
| Some | Reference | Reference | Reference |
| None | 1.25 [0.95–1.65] | 1.42 [1.04–1.94]* | 0.71 [0.36–1.38] |
| **Number of businesses** | | | |
| None | Reference | Reference | Reference |
| Some | 0.79 [0.58–1.08] | 0.85 [0.59–1.21] | 0.65 [0.32–1.30] |
| **Walkability index** | | | |
| Least walkable | 1.21 [0.76–1.93] | 1.37 [0.81–2.33] | 1.11 [0.35–3.58] |
| Below average | 1.36 [0.88–2.10] | 1.44 [0.88–2.35] | 1.29 [0.42–3.97] |
| Above average | 1.46 [0.96–2.24] | 1.30 [0.81–2.09] | 2.28 [0.74–7.06] |
| Most walkable | Reference | Reference | Reference |

Controlled for age at enrollment (continuous), smoking status and pack-year, marital status, employment status, household income, body mass index at enrollment, and family history of prostate cancer. [a]No fast-food restaurant and other restaurants; [b]No fast-food restaurant and retail food; nSES, neighborhood socioeconomic status. BMI, body mass index; * statistically significant (p-value <0.05).

pre-PSA era (31, 32). However, the limited number of deaths from prostate cancer within our study resulted in statistical underpowering, restricting our ability to assess variations among different racial groups. Furthermore, our analysis did not reveal significant associations between most of the neighborhood-built environment attributes (except for the restaurant environment index) and prostate cancer mortality, indicating the need for further research that includes larger studies linking mortality, SDOH, and neighborhood-level data.

This study had several notable strengths. The SCCS is a large prospective cohort study. More than two-thirds of the participants were Black men, and White men in the study had overlapping SES levels. Individual-level data on prostate cancer risk factors were available, enabling an investigation into the effect of neighborhood obesogenic factors while accounting for individual-level factors. We also included multiple indices related to the built environment that are potentially influential in obesity. This study was limited by the lower number of mortality outcomes or participants with tumor pathology available. Moreover, underweight participants were few and were excluded from our stratified analysis due to concerns about inadequate statistical power. Additionally, we were unable to gather information regarding certain neighborhood attributes, such as street connectivity and traffic density, which would be of additional interest. All neighborhood obesogenic factors were gathered based on the participants' addresses provided at the study baseline, and some participants likely moved to a higher- or

TABLE 8 Neighborhood obesogenic factors and incidence of prostate cancer mortality, the Southern Community Cohort Study, 2002 to 2018.

| Variables | Deaths | aHR (95%CI) |
| --- | --- | --- |
| nSES | | |
| Quintile 5 (high) | 32 | Reference |
| Quintile 4 | 22 | 1.78 [0.78–4.09] |
| Quintile 3 | 15 | 1.39 [0.58–3.31] |
| Quintile 2 | 18 | 2.28 [1.01–5.12]* |
| Quantile 1 (low) | 11 | 3.45 [1.54–7.64]* |
| Restaurant environment Index | | |
| None[a] | 68 | Reference |
| Quintile 1 | 14 | 5.12 [1.57–16.67]* |
| Quintile 2 | NR | 1.29 [0.13–12.68] |
| Quintile 3 | NR | 2.71 [0.53–13.81] |
| No restaurant | 11 | 1.46 [0.46–4.63] |
| Retail food environment Index | | |
| None[b] | 49 | Reference |
| Quintile 1 | NR | 0.41 [0.08–2.18] |
| Quintile 2 | NR | 0.35 [0.07–1.74] |
| Quintile 3 | NR | 0.87 [0.19–4.03] |
| No retail food | 40 | 0.79 [0.42–1.51] |
| Number of parks | | |
| Some | 35 | Reference |
| None | 63 | 1.71 [0.63–4.62] |
| Number of recreation facilities | | |
| Some | 33 | Reference |
| None | 65 | 0.79 [0.42–1.50] |
| Number of businesses | | |
| None | 82 | Reference |
| Some | 16 | 1.77 [0.90–3.48] |
| Walkability index | | |
| Least walkable | 26 | 3.45 [1.22–9.78]* |
| Below average | 35 | 1.89 [0.73–4.85] |
| Above average | 29 | 1.65 [0.67–4.04] |
| Most walkable | NR | Reference |
| PSA testing | | |
| Yes | 63 | 0.46 [0.28–0.75]* |
| No or unknown | 35 | Reference |

Controlled for age at enrollment, smoking status and pack-year, marital status, employment status, household income, body mass index at enrollment, family history of prostate cancer, PSA testing, cancer stage, and total sit hours; nSES, neighborhood socioeconomic status. [a]No fast-food restaurant and other restaurants; [b]No fast-food restaurant and retail food. NR, not reportable; * statistically significant (p-value <0.05).

lower-nSES area prior to study entry or during the follow-up period. Additionally, races other than Black and White constituting less than 1% of the SCCS study population were excluded from the analysis, and thus, our results may not generalize across all races/ethnicities.

In conclusion, we found lower nSES to be associated with a higher prostate cancer mortality overall, and a higher prostate cancer risk among Black men. This is consistent with our previous studies (33–37) on the impact of social determinants of health, including neighborhood characteristics, on a range of various chronic and non-chronic conditions. We also found areas with fewer healthy food choices to be associated with prostate cancer among Black men. Results highlight the effects of neighborhood-level risk factors and the possible impact that public health policies could have on prostate cancer outcomes in lower-income areas. Overall, findings represent the need for further exploration of these dynamic associations between SDoH and health outcomes to further reduce suffering for the most vulnerable communities. Our study contributes to the increasing need for contextual evidence emphasizing the importance of examining how neighborhood-built environments impact prostate cancer screening, risks, and mortality.

# Data availability statement

The data analyzed in this study is subject to the following licenses/restrictions: These data can be available in accordance with the data access guidelines established by the Southern Community Cohort Study. Requests to access these datasets should be directed to https://ors.southerncommunitystudy.org.

# Ethics statement

The SCCS received ethical approval from the institutional review boards at Vanderbilt University and Meharry Medical College. All study participants provided written, informed consent prior to participation. In addition, this study was reviewed by the Institutional Review Board of the University of Tennessee Health Science Center for analysis without personal identifiers and was granted a waiver.

# Author contributions

FK: Conceptualization, Data curation, Formal analysis, Investigation, Methodology, Project administration, Resources, Writing – original draft, Writing – review & editing. JF: Conceptualization, Data curation, Methodology, Supervision, Writing – original draft, Writing – review & editing. SH: Data curation, Writing – review & editing. BW: Conceptualization, Project administration, Writing – review & editing. MS:

Conceptualization, Data curation, Methodology, Writing – review & editing. AS: Conceptualization, Data curation, Formal analysis, Funding acquisition, Investigation, Methodology, Project administration, Supervision, Writing – review & editing.

# Funding

The author(s) declare that financial support was received for the research, authorship, and/or publication of this article. This study is partially supported by Grant# 1R37CA234119-01A1 from the National Cancer Institute (NCI). Research reported in this publication was supported by the National Cancer Institute of the National Institutes of Health under Award Number U01CA202979. SCCS data collection was performed by the Survey and Biospecimen Shared Resource, which is supported in part by the Vanderbilt-Ingram Cancer Center (P30 CA68485).

# Acknowledgments

The authors would like to thank the Southern Community Cohort Study team for providing the data and reviewing the manuscript. In addition, we would also like to thank the University of Tennessee Health Science Center (UTHSC)—Oak Ridge National Laboratory (ORNL) Center for Biomedical Informatics for providing the resources needed for conducting this research.

# Conflict of interest

The authors declare that the research was conducted in the absence of any commercial or financial relationships that could be construed as a potential conflict of interest.

# Publisher's note

All claims expressed in this article are solely those of the authors and do not necessarily represent those of their affiliated organizations, or those of the publisher, the editors and the reviewers. Any product that may be evaluated in this article, or claim that may be made by its manufacturer, is not guaranteed or endorsed by the publisher.

# Author disclaimer

The content is solely the responsibility of the authors and does not necessarily represent the official views of the National Institutes of Health.

# References

1. Siegel RL, Miller KD, Jemal A. Cancer statistics, 2019. *CA Cancer J Clin*. (2019) 69:7–34. doi: 10.3322/caac.21551

2. Krstev S, Knutsson A. Occupational risk factors for prostate cancer: A meta-analysis. *J Cancer Prev*. (2019) 24:91–111. doi: 10.15430/JCP.2019.24.2.91

3. Bratt O, Drevin L, Akre O, Garmo H, Stattin P. Family history and probability of prostate cancer, differentiated by risk category: A nationwide population-based study. *JNCI J Natl Cancer Inst*. (2016) 108:1–7. doi: 10.1093/jnci/djw110

4. Page EC, Bancroft EK, Brook MN, Assel M, Battat MHA, Thomas S, et al. Interim results from the IMPACT study: evidence for prostate-specific antigen screening in BRCA2 mutation carriers. *Eur Urol*. (2019) 76:831–42. doi: 10.1016/j.eururo.2019.08.019

5. Solar O, Irwin A. *A conceptual framework for action on the social determinants of health: Social determinants of health discussion paper 2*. Geneva, Switzerland: World Health Organization (2010).

6. Coughlin SS. A review of social determinants of prostate cancer risk, stage, and survival. *Prostate Int*. (2020) 8:49–54. doi: 10.1016/j.prnil.2019.08.001

7. Cheng I, Witte JS, McClure LA, Shema SJ, Cockburn MG, John EM, et al. Socioeconomic status and prostate cancer incidence and mortality rates among the diverse population of California. *Cancer Causes Control*. (2009) 20:1431–40. doi: 10.1007/s10552-009-9369-0

8. Ellis L, Canchola AJ, Spiegel D, Ladabaum U, Haile R, Gomez SL. Racial and ethnic disparities in cancer survival: the contribution of tumor, sociodemographic, institutional, and neighborhood characteristics. *J Of Clin Oncol*. (2018) 36:25–33. doi: 10.1200/JCO.2017.74.2049

9. Vidal AC, Oyekunle T, Howard LE, De Hoedt AM, Kane CJ, Terris MK, et al. et al: Obesity, race, and long-term prostate cancer outcomes. *Cancer*. (2020) 126:3733–41. doi: 10.1002/cncr.32906

10. Freedland S, Giovannucci E, Platz E. Are findings from studies of obesity and prostate cancer really in conflict? *Cancer Causes Control*. (2006) 17:5–9. doi: 10.1007/s10552-005-0378-3

11. Fowke JH, Motley SS, Concepcion RS, Penson DF, Barocas DA. Obesity, body composition, and prostate cancer. *BMC Cancer*. (2012) 12:23. doi: 10.1186/1471-2407-12-23

12. Lin X, Li H. Obesity: epidemiology, pathophysiology, and therapeutics. *Front Endocrinol*. (2021) 12. doi: 10.3389/fendo.2021.706978

13. Javed Z, Valero-Elizondo J, Maqsood MH, Mahajan S, Taha MB, Patel KV, et al. Social determinants of health and obesity: Findings from a national study of US adults. *Obesity*. (2022) 30:491–502. doi: 10.1002/oby.23336

14. DeRouen MC, Tao L, Shariff-Marco S, Yang J, Shvetsov YB, Park SY, et al. Neighborhood obesogenic environment and risk of prostate cancer: the multiethnic cohort. *Cancer Epidemiol Biomarkers Prev*. (2022) 31:972–81. doi: 10.1158/1055-9965.EPI-21-1185

15. Signorello LB, Hargreaves MK, Blot WJ. The southern community cohort study: investigating health disparities. *J Health Care Poor Underserved*. (2010) 21:26–37. doi: 10.1353/hpu.0.0245

16. Signorello LB, Hargreaves MK, Steinwandel MD, Zheng W, Cai Q, Schlundt DG, et al. Southern community cohort study: establishing a cohort to investigate health disparities. *J Natl Med Assoc*. (2005) 97:972–9.

17. Yost K, Perkins C, Cohen R, Morris C, Wright W. Socioeconomic status and breast cancer incidence in California for different race/ethnic groups. *Cancer Causes Control*. (2001) 12:703–11. doi: 10.1023/A:1011240019516

18. Conroy S, Clarke CA, Yang J, Shariff-Marco S, Shvetsov YB, Park S-Y, et al. et al: Contextual impact of neighborhood obesogenic factors on postmenopausal breast cancer: The Multiethnic Cohort. *Cancer Epidemiol Biomarkers Prev*. (2017) 26:480–9. doi: 10.1158/1055-9965.EPI-16-0941

19. United States Enviromental Protection Agency (EPA). *National Walkability Index Methodology and User Guide*. (2021). Available online at: https://www.epa.gov/smartgrowth/national-walkability-index-user-guide-and-methodology.

20. Kilpeläinen TP, Talala K, Raitanen J, Taari K, Kujala P, Tammela TLJ, et al. Prostate cancer and socioeconomic status in the finnish randomized study of screening for prostate cancer. *Am J Epidemiol*. (2016) 184:720–31. doi: 10.1093/aje/kww084

21. Vickers AJ, Mahal B, Ogunwobi OO. Racism does not cause prostate cancer, it causes prostate cancer death. *J Clin Oncol*. (2023) 41:2151–4. doi: 10.1200/JCO.22.02203

22. Lillard JW, Moses KA, Mahal BA, George DJ. Racial disparities in Black men with prostate cancer: A literature review. *Cancer*. (2022) 128:3787–95. doi: 10.1002/cncr.34433

23. Leapman MS, Dinan M, Pasha S, Long J, Washington SL, Ma X, et al. Mediators of racial disparity in the use of prostate magnetic resonance imaging among patients with prostate cancer. *JAMA Oncol*. (2022) 8:687–96. doi: 10.1001/jamaoncol.2021.8116

24. Perry MJ, Arrington S, Freisthler MS, Ibe IN, McCray NL, Neumann LM, et al. Pervasive structural racism in environmental epidemiology. *Environ Health*. (2021) 20:1–21. doi: 10.1186/s12940-021-00801-3

25. Churchwell K, Elkind CMSV, Benjamin RM, Carson AP, Chang EK, Lawrence W, et al. et al: call to action: structural racism as a fundamental driver of health disparities: A presidential advisory from the american heart association. *Circulation*. (2020) 142:e454–68. doi: 10.1161/CIR.0000000000000936

26. Cooksey-Stowers K, Schwartz MB, Brownell KD. Food swamps predict obesity rates better than food deserts in the United States. *Int J Environ Res Public Health*. (2017) 14:1-20. doi: 10.3390/ijerph14111366

27. Bevel MS, Tsai M-H, Parham A, Andrzejak SE, Jones S, Moore JX. Association of food deserts and food swamps with obesity-related cancer mortality in the US. *JAMA Oncol*. (2023) 9:909–16. doi: 10.1001/jamaoncol.2023.0634

28. Vince RA, Jiang R, Bank M, Quarles J, Patel M, Sun Y, et al. evaluation of social determinants of health and prostate cancer outcomes among black and white patients A systematic review and meta-analysis. *JAMA Network Open*. (2023) 6:1–13. doi: 10.1001/jamanetworkopen.2022.50416

29. DeRouen MC, Schupp CW, Koo J, Yang J, Hertz A, Shariff-Marco S, et al. et al: Impact of individual and neighborhood factors on disparities in prostate cancer survival. *Cancer Epidemiol*. (2018) 53:1–11. doi: 10.1016/j.canep.2018.01.003

30. Robinson JRM, Beebe-Dimmer JL, Schwartz AG, Ruterbusch JJ, Baird TE, Pandolfi SS, et al. Neighborhood walkability and body mass index in African American cancer survivors: The Detroit Research on Cancer Survivors Study. *Cancer*. (2021) 127:4687–93. doi: 10.1002/cncr.33869

31. Gulati R, Nyame YA, Lange JM, Shoag JE, Tsodikov A, Etzioni R. Racial disparities in prostate cancer mortality: a model-based decomposition of contributing factors. *J Natl Cancer Institute Monogr*. (2023) 62):212–8. doi: 10.1093/jncimonographs/lgad018

32. Kaur D, Ulloa-Pérez E, Gulati R, Etzioni R. Racial disparities in prostate cancer survival in a screened population: Reality versus artifact. *Cancer*. (2018) 124:1752–9. doi: 10.1002/cncr.31253

33. Brakefield WS, Olusanya OA, Shaban-Nejad A. Association between neighborhood factors and adult obesity in shelby county, tennessee: geospatial machine learning approach. *JMIR Public Health And Surveillance*. (2022) 8:e37039. doi: 10.2196/37039

34. Shin EK, Kwon Y, Shaban-Nejad A. Geo-clustered chronic affinity: pathways from socio-economic disadvantages to health disparities. *JAMIA Open*. (2019) 2:317–22. doi: 10.1093/jamiaopen/ooz029

35. Shin EK, Shaban-Nejad A. Urban decay and pediatric asthma prevalence in memphis, tennessee: urban data integration for efficient population health surveillance. *IEEE Access*. (2018) 6:46281–9. doi: 10.1109/ACCESS.2018.2866069

36. Shin EK, Mahajan R, Akbilgic O, Shaban-Nejad A. Sociomarkers and biomarkers: predictive modeling in identifying pediatric asthma patients at risk of hospital revisits. *NPJ Digital Med*. (2018) 1:1-5. doi: 10.1038/s41746-018-0056-y

37. Brakefield WS, Olusanya OA, White B, Shaban-Nejad A. Social determinants and indicators of COVID-19 among marginalized communities: A scientific review and call to action for pandemic response and recovery. *Disaster Med Public Health Preparedness*. (2023) 17:e193. doi: 10.1017/dmp.2022.104